# Hyper-relativistic mechanics and superluminal particles

## Yu.I. Bogdanov[1], A.Yu. Bogdanov

Recent experiments by OPERA with high energy neutrinos, as well as astrophysics observation data, may possibly prove violations of underlying principles of special relativity theory.

This paper attempts to present an elementary modification of relativistic mechanics that is consistent both with the principles of mechanics and with Dirac's approach to derivation of relativistic quantum equations. Our proposed hyper-relativistic model is based on modified dispersion relations between energy and momentum of a particle.

Predictions of the new theory significantly differ from the standard model, as the former implies large Lorentz gamma-factors (ratio of particle energy to its mass). First of all, we study model relationships that describe hypothetical motion of superluminal neutrinos. Next, we analyze characteristics of Cherenkov radiation of photons and non-zero mass particles in vacuum. Afterwards, we derive generalized Lorentz transformations for a hyper-relativistic case, resulting in a radical change in the law of composition of velocities and particle kinematics. Finally, we study a hyper-relativistic version of Dirac equation and some of its properties.

In present paper we attempted to use plain language to make it accessible not only to scientists but to undergraduate students as well.

**Introduction**

Lorentz invariance is in the heart of special relativity theory and it is a very precise form of symmetry in the Nature. That symmetry is regarded as one of the main attributes of modern elementary particle physics, as well as general description of the world and rightly so.

Recent experiments with neutrinos conducted by OPERA hinted at the possibility of superluminal motion of particles [1]. These results inspired development of new theories that described phenomena beyond the scope of special relativity theory. Previously, such works had usually appeared in fields of quantum gravitation theory and in connection with some astrophysics observations [2,3]. Among such observations let us highlight a study of GZK cutoff (Greisen, Zatsepin, Kuzmin) [4,5]. The described effect is that the energy of cosmic rays (mainly protons) that can reach the Earth cannot exceed the value of around $5 \cdot 10^{19} \, eV$. This is because protons lose energy during interaction with relict microwave radiation, which leads to pi-meson (pions) production. Calculation of GZK-cutoff is essentially based on assumption of validity of special relativity theory for ultra-high energy particles. Existence of registered particles with energy on the order of $10^{20} \, eV$ and higher presents a certain problem for contemporary astrophysics. One of the possible solutions to the problem is to search for violation of Lorentz invariance. In that case the threshold of the reaction of pion photoproduction could shift to higher energy levels that would allow particles with energies higher than GZK-cutoff reach the Earth [6].

Traditional approach to description of violation of Lorentz invariance [2] is to replace the standard dispersion relation $E^2 = p^2 + m^2$ with its more general form $E^2 = F(p,m)$.

[1] Institute of Physics and Technology, Russian Academy of Sciences, Moscow, 117218 Russia, e-mail: bogdanov@ftian.ru

In the traditional approach it is natural to decompose function $F(p,m)$ into its Taylor series around point $p=0$ (rotational symmetry is also assumed).

$$E^2 = m^2 + p^2 + F^{(1)}|p| + F^{(2)}p^2 + F^{(3)}|p|^3 + ... \quad (I.1)$$

Our approach is based on a significantly different (and more simple) form of dispersion relation modification

$$F(E) - p^2 = m^2 \quad (I.2)$$

Here function $F(E)$ has the form $F(E) = E^2$ for Lorentz standard model and is more complex for a hyper-relativistic case when special relativity theory is violated. It appears that such form of dispersion relation is the most natural from standpoint of principles of mechanics. It also wonderfully agrees with Dirac's approach to derivation of relativistic quantum equations. The actual form of function $F(E)$ should be chosen based on phenomenological considerations (e.g. experiments with neutrinos and astrophysical observations).

Our approach to the considered problem is based on a genesis of principal conceptions of mechanics. Three different stages of the genesis are related to the three forms of mechanics– Newtonian, Relativistic and Hyper-relativistic.

Newtonian mechanics encompasses classical notions of time and space, the classical law of composition of velocities, Galilean transformations etc. Laws of relativistic mechanics imply special relativity theory, relativistic law of composition of velocities, Lorentz transformations etc. It appears, though, that classical and standard relativistic mechanics are just particular cases of a more general framework of hyper-relativistic mechanics. That formalism is based on laws of energy and momentum conservation, as well as energy-momentum invariants that naturally arise from description of mechanical systems. It is remarkable that the difference between those three types of mechanics is explained by different definitions of momentum.

As particle energy grows we move from non-relativistic to relativistic energy region, followed by ultra-relativistic and finally by hyper-relativistic field. Principles of hyper-relativistic dynamics significantly change notions of special relativity theory, resulting in a gamma-factor equal to or larger than so-called critical gamma-factor.

It is well known that relativistic laws significantly amend kinematic relationships of classical mechanics, such as the law of composition of velocities, coordinate transformations etc. It appears that hyper-relativistic mechanics calls for an even greater and more radical abandonment of universal kinematic relations as such.

In the hyper-relativistic framework the same system can be described by different kinematic variables. This implies existence of principle of relativity (in some generalized hyper-relativistic sense). We can define hyper-relativistic laws of velocity composition, generalized Lorentz transformations etc. Note, however, that in this case transformation from one set of kinematic variables to another depends not only on velocities, but also on mass of the particles. This phenomenon crucially changes our traditional understanding of kinematics as a field of mechanics that studies motion of particles independent from their mass.

Kinematics then becomes a secondary tool for describing dynamics of a system. It can only be applied in narrow cases where we fix means of generation and registration of particles and their states. Therefore, strictly speaking, in hyper-relativistic mechanics we can no longer apply the traditional notion of kinematics as a universal tool independent from dynamics.

Finally, note that during transition from classical mechanics to special relativity theory, we could preserve the traditional understanding of kinematics by introducing such notions as relativity of simultaneity, time dilation, Lorentz contraction etc. As the result, we had to move from the Newtonian concept of absolute time and space to relative space-time in the theory of relativity. In hyper-relativistic mechanics space-time becomes an auxiliary concept dependent on the mass, which makes it receptive to problems of theory of gravity. At the same time, dependence of results of hyper-relativistic mechanics on observation conditions draws it together to quantum mechanics.

Present paper is structured as follows.

In Section 1 we consider formalism of hyper-relativistic mechanics. As the main postulates we use the principal law of dynamics (Newton's second law), the law of conservation of energy, and the definition of momentum. Three main forms of mechanics – classical, relativistic and hyper-relativistic are studied. We analyze the relation between energy and momentum for particles and their systems. Finally, we apply the notion of energy-momentum invariants to construct the hyper-relativistic law of velocity composition, as well as an analog of Lorentz transformations.

In Section 2 we present some particular models of hyper-relativistic dynamics. We mostly consider a power model with gradual hyper-relativistic adjustment with the growth of energy. Next we study some modifications of the proposed model as well as a number of other elementary models. In light of experiments by OPERA with neutrinos, we construct curves for possible dependence of superluminal velocity of a particle on its energy.

In Section 3 we include some numerical calculations in hyper-relativistic framework. We consider examples related to composition of motion as well as hyper-relativistic Cherenkov radiation of photons and non-zero mass particles in vacuum.

Section 4 is concerned with derivation of hyper-relativistic Dirac equation. We describe application of Dirac's ideology to hyper-relativistic dispersion relation that is based on square root extraction. Afterwards, we study free hyper-relativistic motion of a Dirac particle, as well as effect of an external electromagnetic field.

Finally, a summary of main results of the work is presented in Section 5.

**1. Hyper-relativistic dynamics "for dummies"**
**1.1. Principal equations**

The main parameters that describe motion of a mass particle in mechanics are momentum (the amount of motion), force that defines interaction as the reason for change in momentum, and energy that is changed by acts of forces upon the particle. It is remarkable that the difference between the three types of mechanics (Newtonian, relativistic and hyper-relativistic) is only due to definition of momentum.

Let us postulate that the principal law of motion (Newton's second law), as well as the law of conservation of energy are unchanged.

$$\frac{d}{dt}\vec{p} = \vec{F} \qquad (1.1.1)$$

$$\frac{dE}{dt} = \vec{v} \cdot \vec{F} \qquad (1.1.2)$$

Equation (1.1.1) shows that the rate of change of momentum is the force acting upon the particle. Likewise, equation (1.1.2) shows that the rate of change of energy is the work in unit time, i.e. power, equal to scalar product of force to velocity.

The difference between modified and standard dynamics is due to definition of momentum of a particle, which now becomes equal to:

$$\vec{p} = f(E) \cdot \vec{v}, \qquad (1.1.3)$$

where $f(E)$ is some function of energy.

In classical Newtonian mechanics, function $f(E)$ is constant (and equal to the mass), while in special relativity theory $f(E) = E$. In the hyper-relativistic framework we assume that $f(E)$ is some arbitrary function that is close to $E$ for low energy levels and is significantly non-linear for extremely high energies.

Let us call $f(E)$ - the hyper-relativistic energy function, while its particular form $f(E) = E$ - the standard relativistic (Lorentz) energy function.

## 1.2. Relationship between energy and momentum

From the law of energy conservation (1.1.2), definition of momentum (1.1.3) and the principal law of dynamics (1.1.1), we can derive the following string of identities:

$$\frac{dE}{dt} = \vec{v} \cdot \vec{F} = \frac{\vec{p}}{f(E)} \frac{d\vec{p}}{dt}, \qquad (1.2.1)$$

which leads us to:

$$\int f(E) dE = \int \vec{p} \, d\vec{p} = \int d\frac{p^2}{2} \qquad (1.2.2)$$

Let us analyze the results that equation (1.2.2) produces in the three forms of mechanics.

In Newtonian mechanics $f(E) = m$, which leads to the following dispersion relationship:

$$E - \frac{p^2}{2m} = const \qquad (1.2.3)$$

The constant of integration in (1.2.3) can be calculated by considering a static particle. Let $E_0$ be the rest energy (intrinsic energy of the particle). Then the dispersion relationship in Newtonian mechanics takes the form:

$$E - \frac{p^2}{2m} = E_0 \qquad (1.2.4)$$

In relativistic mechanics momentum is defined by function $f(E) = E$. Then we can derive the following dispersion relationship:

$$E^2 - p^2 = E_0^2 \qquad (1.2.5)$$

From (1.2.5) and (1.1.3) we can calculate the following dependence of energy on velocity:

$$E = \frac{E_0}{\sqrt{1-v^2}} \qquad (1.2.6)$$

Note that we work in a system where the speed of light is normalized ($c=1$)

For low velocities $v \ll 1$ the following approximation holds:

$$E \approx E_0 + \frac{E_0 v^2}{2} \qquad (1.2.7)$$

Furthermore, we need to condition that the derived relationship matches with the equation for kinetic energy in Newtonian mechanics (so called correspondence principle). This leads us to the famous identity correspondence between mass of a particle and its intrinsic energy $E_0$.

$$E_0 = m \qquad (1.2.8)$$

Finally, we rewrite equation (1.2.5) for dispersion relation in relativistic mechanics in the form:

$$E^2 - p^2 = m^2 \qquad (1.2.9)$$

Hyper-relativistic mechanics in general corresponds to some arbitrary function $f(E)$. The only condition we must impose is its approximate coincidence with standard relativistic relation $f(E) = E$ for reasonably small energy levels.

In this case the dispersion relation that follows from (1.2.2) has the form:

$$F(E) - p^2 = const,$$

where the introduced function $F(E)$ is related to function $f(E)$ as:

$$f(E) = \frac{1}{2}\frac{dF}{dE} \qquad (1.2.10)$$

The very same correspondence principle conditions that for low energies we have: $F(E) = E^2$ and $E^2 - p^2 = m^2$. Then, for $p = 0$ we get $E = m$ and $F(m) = m^2$.

Therefore, the general dispersion relation should have the form:

$$F(E) - p^2 = m^2, \qquad (1.2.11)$$

It follows that in all three cases dispersion relations are defined by the same invariant–mass of the particle $m$. In different experiment settings a certain particle would clearly have different values of energy and momentum, but they are interconnected by the dispersion relation, i.e. a energy-momentum invariant defined by the mass. Let us call equation (1.2.11) hyper-relativistic invariant, while standard relation $E^2 - p^2 = m^2$ - relativistic invariant.

### 1.3. Relation between energy and momentum in dimensionless coordinates

Let us introduce dimensionless energy and momentum

$$\gamma = \frac{E}{m} \qquad (1.3.1)$$

$$\vec{\delta} = \frac{\vec{p}}{m} \qquad (1.3.2)$$

In new coordinates the hyper-relativistic invariant will have the form:

$$F_0(\gamma) - \delta^2 = 1 \qquad (1.3.3)$$

Here we also introduced the following dimensionless energy function

$$F_0(\gamma) = \frac{F(E)}{m^2} \qquad (1.3.4)$$

Let

$$f_0(\gamma) = \frac{1}{2}\frac{dF_0}{d\gamma}, \qquad (1.3.5)$$

Then

$$\vec{\delta} = f_0(\gamma)\vec{v} \qquad (1.3.6)$$

Relationship between velocity and gamma-factor is given by the following equation

$$v = \frac{\sqrt{F_0(\gamma) - 1}}{f_0(\gamma)} \qquad (1.3.7)$$

For standard model $F_0(\gamma) = \gamma^2$, $f_0(\gamma) = \gamma$, which leads to the standard relation between velocity and gamma-factor:

$$v = \frac{\sqrt{\gamma^2 - 1}}{\gamma}, \qquad (1.3.8)$$

or

$$\gamma = \frac{1}{\sqrt{1 - v^2}} \qquad (1.3.9)$$

Apart from the dispersion function $F(E)$ it is useful to introduce its square root

$$G(E) = \sqrt{F(E)} \qquad (1.3.10)$$

Then dispersion equation (1.2.11) takes the form:

$$G^2(E) - p^2 = m^2, \qquad (1.3.11)$$

Let us also introduce dimensionless function

$$G_0(\gamma) = \sqrt{F_0(\gamma)} \qquad (1.3.12)$$

Then:

$$G_0^2(\gamma) - \delta^2 = 1 \qquad (1.3.13)$$

Equations (1.3.11) and (1.3.13) demonstrate an obvious similarity to pseudo-Euclidean space. However, that analogy is incomplete. For instance, for a system of two free particles the energy is additive, thus $E = E_1 + E_2$, but more generally function $G(E)$ is non-additive since $G(E_1 + E_2) \neq G(E_1) + G(E_2)$. The exception, when $G(E_1 + E_2) = G(E_1) + G(E_2)$, only holds for the standard Lorentz case $G(E) = E$.

**1.4. Description of system of particles**

Let total energy and momentum of a system be a composition from individual particles:

$$E = E_1 + E_2 + \ldots + E_n \qquad (1.4.1)$$

$$\vec{p} = \vec{p}_1 + \vec{p}_2 + \ldots + \vec{p}_n \qquad (1.4.2)$$

Energy-momentum invariant of the system as a whole is

$$F(E) - p^2 = F(E_1 + \ldots + E_n) - (\vec{p}_1 + \ldots + \vec{p}_n)^2 = M^2 \qquad (1.4.3)$$

This equation defines mass $M$ of the system as a whole. A typical application is concerned with the problem of scattering (collision). We suppose that particles are in free motion both at the input and the output. During collision the energies, momenta and types of particles can change, but total energy, momentum and, therefore, mass $M$ remain the same. For instance, the total mass of a system may define a threshold for certain reactions. Note also that our particles move uniformly straight forward, so our description refers to an inertial frame of reference.

Let us study the system of two particles in more detail.

In dimensionless coordinates the laws of conservation of energy and momentum have the form:

$$m_1 \gamma_1 + m_2 \gamma_2 = M \gamma \qquad (1.4.4)$$

$$m_1 \vec{\delta}_1 + m_2 \vec{\delta}_2 = M \vec{\delta} \qquad (1.4.5)$$

Then

$$\frac{|\vec{\delta}|}{\gamma} = \frac{|m_1 \vec{\delta}_1 + m_2 \vec{\delta}_2|}{m_1 \gamma_1 + m_2 \gamma_2} \qquad (1.4.6)$$

and

$$F_0(\gamma) - \left( \frac{m_1 \vec{\delta}_1 + m_2 \vec{\delta}_2}{m_1 \gamma_1 + m_2 \gamma_2} \right)^2 \gamma^2 = 1 \qquad (1.4.7)$$

This equation helps us calculate gamma-factor of the system. Then, we can calculate mass and momentum as follows:

$$M = \frac{m_1\gamma_1 + m_2\gamma_2}{\gamma} \quad (1.4.8)$$

$$\vec{\delta} = \frac{m_1\vec{\delta}_1 + m_2\vec{\delta}_2}{M} \quad (1.4.9)$$

Equations (1.4.7) - (1.4.9) can be easily generalized for systems of any number of particles:

$$F_0(\gamma) - \left(\frac{m_1\vec{\delta}_1 + m_2\vec{\delta}_2 + ... + m_n\vec{\delta}_n}{m_1\gamma_1 + m_2\gamma_2 + ... + m_n\gamma_n}\right)^2 \gamma^2 = 1 \quad (1.4.10)$$

$$M = \frac{m_1\gamma_1 + m_2\gamma_2 + ... + m_n\gamma_n}{\gamma} \quad (1.4.11)$$

$$\vec{\delta} = \frac{m_1\vec{\delta}_1 + m_2\vec{\delta}_2 + ... + m_n\vec{\delta}_n}{M} \quad (1.4.12)$$

In a system of $n$ relativistic particles, the particle under number $j = 1,2,...,n$ is characterized by set $(m_j, \gamma_j, \vec{\delta}_j)$, while $F_0(\gamma_j) - \delta_j^2 = 1$. The system as a whole is characterized by set $(M, \gamma, \vec{\delta})$ in accordance with equations (1.4.10)- (1.4.12). In some sense, the whole system can be reduced to a single particle with parameters $(M, \gamma, \vec{\delta})$. More specifically, we can articulate the main geometric property of a system of relativistic particles. Let us divide the system of $n$ particles into two subsystems $A$ and $B$ ($n = n_A + n_B$) and associate particles with parameters $(M_A, \gamma_A, \vec{\delta}_A)$ and $(M_B, \gamma_B, \vec{\delta}_B)$ with those subsystems. We can apply equations (1.4.10)-(1.4.12) to subsystems $A$ and $B$ for that matter. Then calculation of a characteristic set for the original system of $n$ particles $(M, \gamma, \vec{\delta})$ will be equivalent to calculation for a system with just two particles associated with $A$ и $B$.

Note that from (1.2.4) it follows that in non-relativistic case instead of (1.4.3) we will have the following invariant:

$$(E_1 + ... + E_n) - \frac{(\vec{p}_1 + ... + \vec{p}_n)^2}{2(m_1 + ... + m_n)} = E_0 \quad (1.4.13),$$

where $E_0$ is the total rest energy of particles of the system.

Finally, in order to calculate probabilities and cross sections of different processes it is important to define invariant phase volume. Based on energy-momentum invariant (1.2.11), we may define phase volume for a particle as

$$2\delta(F(E) - p^2 - m^2) dE dp_x dp_y dp_z \qquad (1.4.14)$$

Here all integration variables are independent. Therefore, considering delta-function properties during integration and using (1.2.10), we get the following invariant:

$$\frac{dp_x dp_y dp_z}{f(E)} = \frac{d^3 p}{f(E)} \qquad (1.4.15)$$

In (1.4.15) energy and momentum are connected by dispersion relation (1.2.11).
For a system of particles the invariant volume has the form:

$$\frac{d^3 p_1 ... d^3 p_n}{f(E_1)...f(E_n)} \qquad (1.4.16)$$

Finally, for standard Lorentz case we get the well-known result:

$$\frac{d^3 p_1 ... d^3 p_n}{f(E_1)...f(E_n)} = \frac{d^3 p_1 ... d^3 p_n}{E_1...E_n} \qquad (1.4.17)$$

**1.5. Composition of velocities in hyper-relativistic mechanics**

Knowledge of invariant mass $M$ allows one to solve the problem of velocity composition, i.e. move on from the laboratory to some other (active) reference frame. To be sure, clearly we are not referring to a hypothetical physical transition to a relativistic or even superluminal reference frame to conduct real experiments there. What we aim to achieve with the introduction of new kinematic variables is to conduct a complimentary experiment in the actual laboratory reference frame. It is important that this new experiment is as feasible by the laws of Nature as the original one.

Traditional approach to composition of velocities is based on Lorentz coordinate transformations. They can be further used to prove invariance of such dynamic parameters as mass of a system or its subsystem. Nevertheless, there is a reverse approach to the problem. We can postulate an energy-momentum invariant (1.4.3) that defines squared mass of the system and use it to substantiate validity of relativistic law of composition of velocities [7]. Likewise, if we postulate non-relativistic invariant (1.4.13), we can derive the classical law of velocity composition and Galilean transformations.

The advantages of the approach based on energy-momentum invariants become evident during phenomenological construction of hyper-relativistic mechanics. As such, we do not have to come up with explicit and usually artificial a priori models of space-time and can focus instead on developing modified dispersion relations that can be tested experimentally. It appears that in this approach the very existence of universal relativistic kinematics independent from mass of particles can only be accidental. When we move on from the standard law of dispersion, there are significant changes in the laws of kinematics. In fact, the actual traditional notion of a reference frame as a general platform for describing all processes in the Universe loses its physical sense as such.

Let us formulate the main **principle of hyper-relativity** (hyper-relativism) as following: transition from one set of variables (energies and momenta of particles of a system) to another set is permissible, if it preserves the mass of the system as a whole, as well as the mass of its subsystems. Then the new set of variables describes the same feasible experiment in the Nature as the original one did.

Let us study systems of two and three particles in more detail. We start with a two-particle system with respective masses $m_1$ and $m_2$, and we consider a reference frame connected to the first particle. In the new coordinate frame only the second particle is moving, the momentum of the first particle is equal to zero, while its energy is $m_1$. The laws of conservation in this system are the following:

$$m_1 + m_2\gamma'_2 = M\gamma' \qquad (1.5.1)$$

$$m_2\delta'_2 = M\delta' \qquad (1.5.2)$$

We assume that motion of the second particle is along a defined direction in space. In fact, such direction can be chosen arbitrarily by rotation of Cartesian coordinate system.

Next, we shall use dimensionless dispersion relations $F_0(\gamma') - \delta'^2 = 1$ and $F_0(\gamma'_2) - \delta'^2_2 = 1$ for the system as a whole and the second particle respectively, which would give us an equation for the energy of the second particle $\gamma'_2$ in the moving primed coordinate system

$$F_0\left(\frac{m_1 + m_2\gamma'_2}{M}\right) - \left(\frac{m_2}{M}\right)^2 F_0(\gamma'_2) = \frac{M^2 - m_2^2}{M^2} \qquad (1.5.3)$$

As we know the energy $E'_2 = m_2\gamma'_2$ of the particle, we can find its momentum and velocity as:

$$p'_2 = \sqrt{F(E'_2) - m_2^2} \qquad (1.5.4)$$

$$v'_2 = \frac{p'_2}{f(E'_2)} = \frac{\sqrt{F_0(\gamma'_2) - 1}}{f_0(\gamma'_2)} \qquad (1.5.5)$$

An algorithm based on equations (1.5.3)-(1.5.5) defines the actual law of velocity composition.

In general, we definitely cannot limit ourselves to consideration of two-particle systems. Quite often we have to address three— and multi-particle problems such as in muon decay $\mu^- \to e^- + \nu_\mu + \bar{\nu}_e$, as well as hypothetical radiation of electron-positron pairs by a high-energy neutrino, e.g. $\nu_\mu \to \nu_\mu + e^- + e^+$. Note that four-particle and other systems can be studied similarly to the three-particle system.

In the case of three-particle systems we can define the following energy-momentum invariants:

$$s = F(E_1 + E_2 + E_3) - (\vec{p}_1 + \vec{p}_2 + \vec{p}_3)^2 = M^2 \qquad (1.5.6)$$

$$s_{12} = F(E_1 + E_2) - (\vec{p}_1 + \vec{p}_2)^2 \qquad (1.5.7)$$

$$s_{13} = F(E_1 + E_3) - (\vec{p}_1 + \vec{p}_3)^2 \tag{1.5.8}$$

$$s_{23} = F(E_2 + E_3) - (\vec{p}_2 + \vec{p}_3)^2 \tag{1.5.9}$$

The value $\sqrt{s} = M$ defines invariant mass of the whole system, while value $\sqrt{s_{12}}$ defines invariant mass of the subsystem of the first and second particles etc.

Summing up equations (1.5.7)-(1.5.9) and taking into account (1.5.6) and (1.2.11) we get the following:

$$s_{12} + s_{13} + s_{23} = M^2 + m_1^2 + m_2^2 + m_3^2 + [F_{12}(E_1 + E_2) + F_{13}(E_1 + E_3) + F_{23}(E_2 + E_3)] - [F_{123}(E_1 + E_2 + E_3) + F_1(E_1) + F_2(E_2) + F_3(E_3)] \tag{1.5.10}$$

Here we introduced more detailed notations for dispersion functions $F$. For example $F_{123}$ is a dispersion function for the whole system, $F_{12}$ is the dispersion function for the first and second particles' subsystem, while $F_1$ is the dispersion function related to the first particle etc. Such clarification is needed to avoid confusion, as different functions may take different masses as parameters. In cases when the context is clear we shall continue to use the simpler notation.

In the case of a standard Lorentz model, summands in (1.5.10) that include dispersion functions $F$ mutually disappear, which produces a well-known relation in special relativity theory:

$$s_{12} + s_{13} + s_{23} = M^2 + m_1^2 + m_2^2 + m_3^2 \tag{1.5.11}$$

Let us now move to the system connected to the first particle $m_1$. Here $\vec{p}_1' = 0$, $E_1' = m_1$. We shall try to construct kinematics of particles 2 and 3 that would be consistent with invariants (1.5.6)-(1.5.9). We can use a subsystem of particles 1 and 2 and respective invariant $s_{12}$ to calculate energy $E_2'$ and momentum $p_2'$ of the second particle (its motion direction can be chosen arbitrarily). These values can be found from the following system of two equations:

$$s_{12} = F_{12}(m_1 + E_2') - p_2'^2 \tag{1.5.12}$$

$$F_2(E_2') - p_2'^2 = m_2^2 \tag{1.5.13}$$

Therefore:

$$F_{12}(m_1 + E_2') - F_2(E_2') = s_{12} - m_2^2 \tag{1.5.14}$$

If we move on to dimensionless variables, this equation will be equivalent to equation (1.5.3).

Similarly, as explained above, we can use invariant $s_{13}$ to calculate energy $E_3'$ and momentum $p_3'$ of the third particle.

Next, to find the angle $\theta_{23}'$ between directions of motion of the second and third particles in the primed coordinate system, we can use either invariant $s_{23}$ or invariant $s$. As the result we get:

$$\cos\theta'_{23} = \frac{F_{23}(E'_2 + E'_3) - p'^2_2 - p'^2_3 - s_{23}}{2p'_2 p'_3} \qquad (1.5.15)$$

$$\cos\theta'_{23} = \frac{F_{123}(m_1 + E'_2 + E'_3) - p'^2_2 - p'^2_3 - s}{2p'_2 p'_3} \qquad (1.5.16)$$

It is remarkable that the two variants are the same due to condition (1.5.10).

Finally, let us consider an example of a reference frame connected to the mass center (inertia center).

Here instead of (1.5.1) and (1.5.2) we have:

$$m_1 \gamma'_1 + m_2 \gamma'_2 = M \qquad (1.5.17)$$

$$m_1 \delta'_1 + m_2 \delta'_2 = 0 \qquad (1.5.18)$$

The first equation stands for static inertia center ($\gamma' = 1$), while the second one means that total momentum in the system is equal to zero. Note that the direction along which the particles move (in opposite directions) can be chosen arbitrarily.

Then, using dimensionless dispersion relations $F_0(\gamma'_1) - \delta'^2_1 = 1$ and $F_0(\gamma'_2) - \delta'^2_2 = 1$ we derive the following equation for the energy of the second particle $\gamma'_2$ instead of (1.5.3):

$$F_0\left(\frac{M - m_2 \gamma'_2}{m_1}\right) - \left(\frac{m_2}{m_1}\right)^2 F_0(\gamma'_2) = \frac{m_1^2 - m_2^2}{m_1^2} \qquad (1.5.19)$$

After solving this equation, we can easily find all other kinematics variables.

We can also apply the same analysis as above to demonstrate an algorithm for composition of mechanical motion in the general case.

The crucial difference between composition of motion in hyper-relativistic mechanics and special relativity theory is dependence of respective laws of transformation on mass of the particles. For instance, for a two-particle system we can move on from the laboratory coordinate frame which defines velocities $\vec{v}_1$ and $\vec{v}_2$ of the particles to a coordinate system connected to the first particle ($\vec{v}'_1 = 0$). Clearly, in a general case of arbitrary dispersion relation $F(E)$, the result of velocity composition (1.5.5) depends not only on velocities of the particles but also on the mass.

It is remarkable though that in case of standard Lorentz model when $F_0(\gamma) = \gamma^2$, equation (1.5.5) leads to a well-known law of composition of velocities in special relativity theory, which is purely kinematic and does not depend on the mass. In that case, if the second particle chases the first one, then the velocity of the former in the reference frame defined by the latter is:

$$v'_2 = \frac{v_2 - v_1}{1 - v_1 v_2} \qquad (1.5.20)$$

Similarly, if the particles move towards each other:

$$v'_2 = \frac{v_2 + v_1}{1 + v_1 v_2} \qquad (1.5.21)$$

Nevertheless, in the general case of hyper-relativistic models, kinematics becomes less defined. Then, kinematic transformations merely become a narrow tool for description of dynamic systems. In fact, when we describe the same system in different kinematic variables, we have mutually complementary experiments (in meaning attributed by Bohr). As such, when we consider a reference frame where the particles are moving quickly, we have to use a much different registration setting than when considering a frame where the particles are static or move slowly. Therefore, it is impossible to define any universal physical picture of processes in space-time, so that the picture is independent from registration devices and experiment settings, as those differ for various reference frames. We will illustrate this point in the next section.

The classical school of thought acts on the premise that space-time is like a theatre stage. The play is observed by some spectators (either standing still, or moving in a train, on a rocket etc). The spectators do not impact the stage performance or one another. As such, they observe different pictures that can be easily converted by kinematic equations.

It is well-known that such classical notions (so-called Einstein's realism) deeply contradict the foundations of quantum mechanics. Moreover, the example of composition of motion in hyper-relativistic mechanics shows that the Einstein's realism is violated even without direct reference to quantum mechanics.

In hyper-relativistic dynamics, in complete analogy to quantum mechanics, various reference frames characterized by different registration settings, define a set of mutually-complementary experiments. In the framework of standard Lorentz model the experiments can be easily converted into one another and do not possess any extra information as such. This is not the case in hyper-relativistic mechanics. For instance, it may appear that a neutrino is stable in a static or slowly moving reference frame, while it emits Cherenkov radiation in a frame where it moves quickly. In that case, the set of mutually-complementary experiments can become the basis for calculation of non-trivial law of dispersion for the neutrino.

**1.6. Analog of Lorentz transformations in hyper-relativistic mechanics**

Let us consider a two-particle system again. We will describe in detail the algorithm of transition from the laboratory reference frame to the frame connected to the first particle. Let us introduce a parameter equal to the ratio of masses of the particles

$$\mu_{21} = \frac{m_2}{m_1} \qquad (1.6.1)$$

We rewrite (1.4.7) and (1.5.3) as follows:

$$F_0(\gamma) - \left(\frac{\vec{\delta}_1 + \mu_{21}\vec{\delta}_2}{\gamma_1 + \mu_{21}\gamma_2}\right)^2 \gamma^2 = 1 \qquad (1.6.2)$$

$$F_0\left(\frac{\gamma}{(\gamma_1 + \mu_{21}\gamma_2)}(1 + \mu_{21}\gamma'_2)\right) - \left(\frac{\mu_{21}\gamma}{(\gamma_1 + \mu_{21}\gamma_2)}\right)^2 F_0(\gamma'_2) = 1 - \left(\frac{\mu_{21}\gamma}{(\gamma_1 + \mu_{21}\gamma_2)}\right)^2 \qquad (1.6.3)$$

Let $\theta_{21}$ - be the angle between vectors $\vec{\delta}_2$ and $\vec{\delta}_1$. It is defined by a scalar product

of these vectors:

$$\cos\theta_{21} = \frac{\vec{\delta}_1 \vec{\delta}_2}{\delta_1 \delta_2} \qquad (1.6.4)$$

According to (1.3.3), the magnitudes of vectors $\vec{\delta}_2$ and $\vec{\delta}_1$ are defined by the following equations

$$\delta_1 = \sqrt{F_0(\gamma_1) - 1} \qquad (1.6.5)$$

$$\delta_2 = \sqrt{F_0(\gamma_2) - 1} \qquad (1.6.6)$$

Let us further assume that (1.6.5) and (1.6.6) define one-to-one correspondence.

Gamma-factor $\gamma'_2$ plays the role of dimensionless energy in the primed coordinate frame. The combined equations above allow one to derive $\gamma'_2$ as a function of $\gamma_2$, while $\gamma_1$, $\theta_{21}$ and $\mu_{21}$ act as parameters of the transformation. We shall denote the function as $\Gamma$:

$$\gamma'_2 = \Gamma(\gamma_2 | \gamma_1, \theta_{21}, \mu_{21}) \qquad (1.6.7)$$

Then, since we know function $\Gamma$ and taking $\delta'_2 = \sqrt{F_0(\gamma'_2) - 1}$ into account we can define another function $\Delta$, such as:

$$\delta'_2 = \Delta(\delta_2 | \delta_1, \theta_{21}, \mu_{21}) \qquad (1.6.8)$$

The new function describes transformation from dimensionless momentum $\delta_2$ in the laboratory system to dimensionless momentum $\delta'_2$ in the moving system. It is also evident from (1.6.5) that use of parameter $\delta_1$ is equivalent to use of parameter $\gamma_1$. Direction of vector $\vec{\delta}'_2$ can be chosen arbitrary.

These functions $\Gamma$ and $\Delta$ can be calculated numerically for any dispersion law $F_0(\gamma)$. At first stage, we have to calculate $\gamma$. Next, we input the gamma-factor into (1.6.3), solving which we can derive the actual function (1.6.7).

In the particular case of standard dispersion function in special relativity theory $F_0(\gamma) = \gamma^2$ the algorithm above is reduced to usual Lorentz transformations. For example, let us consider Lorentz boost-transformation that corresponds to parallel (collinear) motion of the particles when $\theta_{21} = 0$, or $\theta_{21} = \pi$. Then, standard Lorentz transformation for energy and momentum in our notation will have the following form:

$$\gamma'_2 = \gamma_1 \gamma_2 \mp \delta_1 \delta_2 \qquad (1.6.9)$$

$$\delta'_2 = \gamma_1 \delta_2 \mp \delta_1 \gamma_2 \qquad (1.6.10)$$

Taking into account (1.6.5) and (1.6.6) we obtain explicit elementary expressions for functions $\Gamma$ and $\Delta$:

$$\gamma'_2 = \gamma_1\gamma_2 \mp \sqrt{\gamma_1^2 - 1}\sqrt{\gamma_2^2 - 1} \qquad (1.6.11)$$

$$\delta'_2 = \sqrt{\delta_1^2 + 1}\,\delta_2 \mp \delta_1\sqrt{\delta_2^2 + 1} \qquad (1.6.12)$$

In equations (1.6.9)-(1.6.12) the "minus" sign corresponds to $\theta_{21} = 0$, while the "plus" sign to $\theta_{21} = \pi$.

As the result, we can again observe the difference between Lorentz transformations (1.6.7)-(1.6.8) in hyper-relativistic mechanics and standard relativity theory, i.e. the dependence of transformations on mass of particles. As such, kinematics of the second particle depends on the mass of the first particle, which serves as the body of reference.

### 1.7 Doppler effect in "hyper-particle + photon" system

Let us imagine a hyper-particle with mass $m_1$, gamma-factor $\gamma_1$ and dimensionless momentum $\vec{\delta}_1$, and a photon with energy $\omega_2$ and momentum $\vec{k}_2$ (we assume $\hbar = 1$). We shall treat the photon as a zero-mass particle with ordinary dispersion relation:

$$\left|\vec{k}_2\right| = \omega_2 \qquad (1.7.1)$$

For the mass particle the dispersion law is hyper-relativistic:

$$F_0(\gamma_1) - \delta_1^2 = 1 \qquad (1.7.2)$$

The mass of system "hyper-particle + photon" is defined similar to (1.4.8):

$$M = \frac{m_1\gamma_1 + \omega_2}{\gamma} \qquad (1.7.3),$$

where gamma-factor of the system is defined similar to (1.4.7) by the following equation:

$$F_0(\gamma) - \left(\frac{m_1\vec{\delta}_1 + \vec{k}_2}{m_1\gamma_1 + \omega_2}\right)^2 \gamma^2 = 1 \qquad (1.7.4)$$

Note that in (1.7.2) and (1.7.4) we implicitly assume that the compound system follows the same dispersion law as the original particle but with another mass.

Next, let us move on to the reference frame connected to the hyper-particle. Here the particle is static $\gamma'_1 = 0$, while photon frequency $\omega'_2$ is defined by the following equation that is similar to (1.5.3):

$$F_0\left(\frac{m_1 + \omega'_2}{M}\right) - \left(\frac{\omega'_2}{M}\right)^2 = 1 \qquad (1.7.5)$$

This algorithm of recalculation of frequency $\omega_2$ into frequency $\omega'_2$ defines the actual Doppler effect. When the dispersion relation has the standard Lorentz form, the hyper-relativistic Doppler effect becomes the usual one. For example if the particle and the photon move in parallel with the particle chasing the photon, we get:

$$\omega'_2 = \frac{\omega_2(1 - v_1)}{\sqrt{1 - v_1^2}} \qquad (1.7.6),$$

where $v_1$ -is velocity of the particle in laboratory reference frame.

## 2. Some specific models of hyper-relativistic dynamics.
### 2.1. Power model with slow switch of hyper-relativistic adjustment with growth of energy

In order to devise a feasible model in hyper-relativistic mechanics, one of specific requirements we have to meet is to ensure a very close agreement with special relativity theory across a wide range of energies. Any significant deviations from the laws of special relativity theory can only be observed for extremely large values of Lorentz gamma-factors. As such, we need to be able to perform calculations with high precision even when particle energy changes by magnitude of many orders. To achieve this, it is reasonable to use arbitrary-precision arithmetic algorithms capable of calculating in multi-digit representation of numbers. In present work, we used 128-digit representation to control for stability of results.

The pretext for our construction of the model below were the results of experiments with neutrinos by OPERA. Let us consider the following dispersion function $F(E)$:

$$F(E) = m^{\alpha \tanh\left(\frac{E}{m\gamma_0}\right)} E^{2 - \alpha \tanh\left(\frac{E}{m\gamma_0}\right)} \qquad (2.1.1)$$

Here parameter $\alpha$ describes deviation of the index from the standard value equal to 2. At the same time this deviation only comes into effect for large energy levels, when the gamma-factor is comparable in magnitude or greater than some very large value $\gamma_0$ (critical gamma-factor). For $E \gg m\gamma_0$ the hyperbolic tangent comes close to unity, which corresponds to complete "switch-on" of the hyper-relativistic adjustment.

From (1.2.10) we can derive:

$$f(E) = E \exp\left(\alpha \tanh\left(\frac{E}{m\gamma_0}\right) \ln\left(\frac{m}{E}\right)\right) \left[1 - \frac{\alpha}{2}\left(\tanh\left(\frac{E}{m\gamma_0}\right) + \frac{E}{m\gamma_0} \frac{\ln\left(\frac{E}{m}\right)}{\cosh^2\left(\frac{E}{m\gamma_0}\right)}\right)\right] \qquad (2.1.2)$$

It is convenient to analyze this and similar equations as follows. Firstly, we define energy $E$. Next we calculate momentum and velocity using formulas:

$$p = \sqrt{F(E) - m^2} \qquad (2.1.3)$$

$$v = \frac{p}{f(E)} \qquad (2.1.4)$$

Note that in dimensionless coordinates equation (2.1.1) can be rewritten in a more compact form:

$$F_0(\gamma) = \gamma^2 \gamma^{-\alpha \tanh\left(\frac{\gamma}{\gamma_0}\right)} \qquad (2.1.5)$$

Also, note that the ordinary power model follows from equations (2.1.1) and (2.1.2) in the limit $\gamma_0 \to 0$:

$$F(E) = m^\alpha E^{2-\alpha} \qquad (2.1.6)$$

$$f(E) = \left(1 - \frac{\alpha}{2}\right) m^\alpha E^{1-\alpha} \qquad (2.1.7)$$

Fig. 1 shows a possible dependence of superluminal velocity on energy for a very light particle like a neutrino (parameters are as follows: $m = 0.1 eV$, $\gamma_0 = 1 \cdot 10^{10}$, $\alpha = 1.8 \cdot 10^{-6}$ )

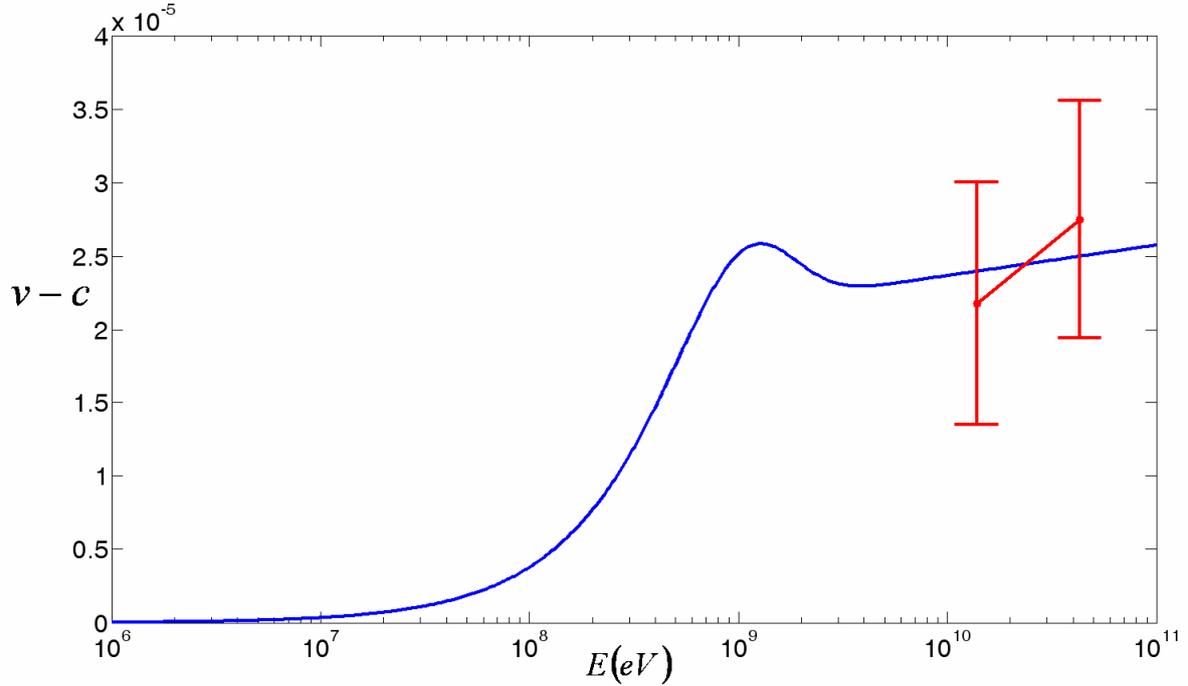

Fig.1. Hypothetical dependence of superluminal velocity on energy for a light particle.

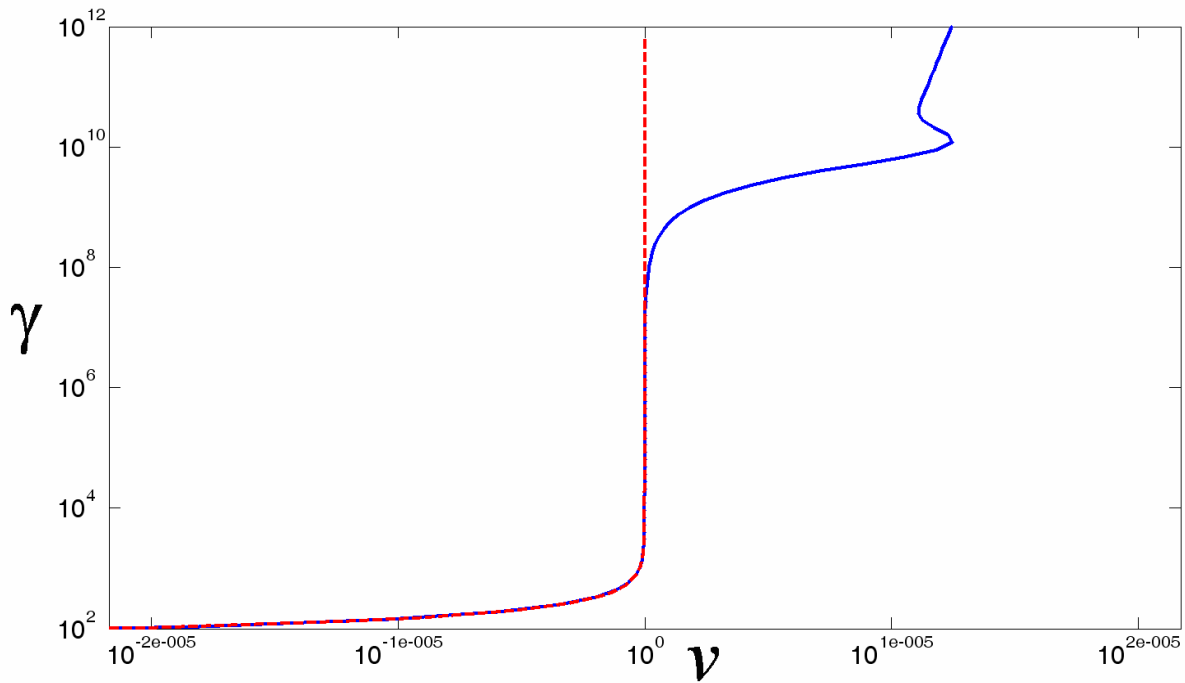

Fig.2 Dependence of gamma-factor on velocity of a particle ( $m = 0.1 eV$, $\gamma_0 = 1 \cdot 10^{10}$, $\alpha = 2 \cdot 10^{-6}$ )

Experimental data from [1] that corresponds to two samples of particles with average energies 13.9 GeV and 42.9 GeV is also shown on the Fig.1.

Finally, the picture on Fig. 1 can be "rotated" to get dependence of gamma-factor on velocity (Fig.2).

Departure of the solid line on Fig. 2 to superluminal field defines the hyper-relativistic anomaly.

**2.2. Modified power model**

According to results by OPERA, anomalous superluminal velocity of neutrino $v - c$ is around $2.5 \cdot 10^{-5}$ for energies on the order of 10-50 GeV. At the same time, the data from supernova SN1987a, shows anomalous velocity of not greater than $2 \cdot 10^{-9}$ for energies on the order of 10 MeV [1,8]. Therefore, the hypothetical anomaly must quickly change depending on the energy in order to be consistent with all available data.

Let us slightly modify our model above to provide for sharper change of energy compared to (2.1.1). To achieve this we shall replace $\dfrac{E}{m\gamma_0}$ with its square in (2.1.1). Therefore:

$$F(E) = m^{\alpha \tanh\left(\left(\frac{E}{m\gamma_0}\right)^2\right)} E^{2 - \alpha \tanh\left(\left(\frac{E}{m\gamma_0}\right)^2\right)} \qquad (2.2.1)$$

We shall also replace (2.1.2) with the following:

$$f(E) = E \exp\left(\alpha \tanh\left(\left(\frac{E}{m\gamma_0}\right)^2\right) \ln\left(\frac{m}{E}\right)\right) \left[1 - \frac{\alpha}{2}\left(\tanh\left(\left(\frac{E}{m\gamma_0}\right)^2\right) + 2\left(\frac{E}{m\gamma_0}\right)^2 \frac{\ln\left(\frac{E}{m}\right)}{\cosh^2\left(\left(\frac{E}{m\gamma_0}\right)^2\right)}\right)\right] \qquad (2.2.2)$$

This model can be further generalized if we consider k-th power instead of square:

$$F(E) = m^{\alpha \tanh\left(\left(\frac{E}{m\gamma_0}\right)^k\right)} E^{2 - \alpha \tanh\left(\left(\frac{E}{m\gamma_0}\right)^k\right)} \qquad (2.2.3)$$

Here the new parameter $k$ describes the speed of "switching-on" of the superluminal anomaly.

Therefore, instead of (2.2.2) we get the following relation:

$$f(E) = E \exp\left(\alpha \tanh\left(\left(\frac{E}{m\gamma_0}\right)^k\right) \ln\left(\frac{m}{E}\right)\right) \left[1 - \frac{\alpha}{2}\left(\tanh\left(\left(\frac{E}{m\gamma_0}\right)^k\right) + k\left(\frac{E}{m\gamma_0}\right)^k \frac{\ln\left(\frac{E}{m}\right)}{\cosh^2\left(\left(\frac{E}{m\gamma_0}\right)^k\right)}\right)\right] \qquad (2.2.4)$$

Fig. 3 is similar to Fig. 1 and shows possible dependence of superluminal velocity of neutrino on its energy (parameter values: $m = 0.1 eV$, $k = 2$, $\gamma_0 = 3 \cdot 10^{10}$, $\alpha = 1.8 \cdot 10^{-6}$).

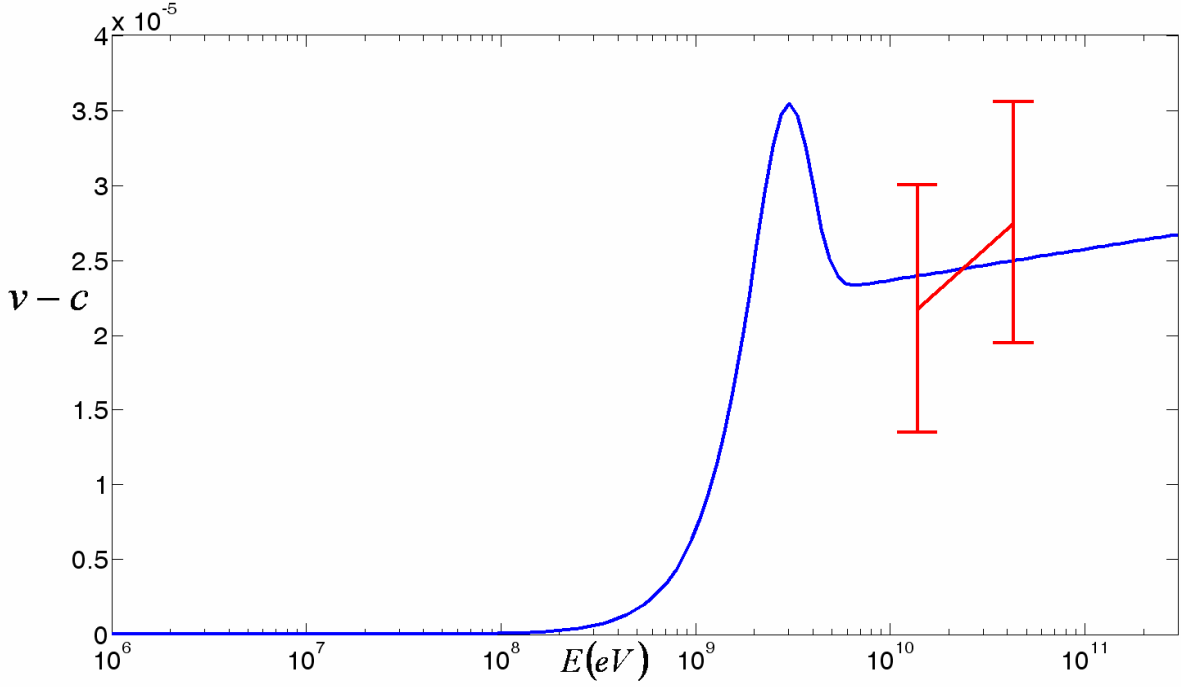

Fig.3. Hypothetical dependence of superluminal speed on energy for neutrino (modified power model)

## 2.3. Renorm transformations (scaling)

Let us consider the following scale transformations in the framework of hyper-relativistic modified power model.

We shall increase significantly the critical gamma-factor $\gamma_0$, ($\gamma_0 \to \gamma_0 K$, where $K$ - is a large multiplier, while the mass of the particle will decrease to the same order $m \to m/K$ so that $\gamma_0 m = const$. Parameter $\alpha$ shall be smoothly scaled down according to the rule: $\alpha \to \eta\alpha$, where $\eta = \dfrac{\log(\gamma)}{\log(\gamma K)}$, $\gamma = \dfrac{E}{m}$, $E$ - is the typical energy level in the problem.

It appears that this transformation does not significantly change ultra-relativistic and hyper-relativistic behavior of the neutrino. At the same time it does radically change its non-relativistic behavior that is however almost undetectable in modern accelerators.

This phenomenon is illustrated on Fig. 4 that is analogous to Fig.1 and Fig.3.

Here we used a modified power model (2.2.3) with $k = 2$. Parameters of the models are as follows:

Model 1: $\alpha = 1.8 \cdot 10^{-6}$, $m = 0.1 eV$, $\gamma_0 = 3 \cdot 10^{10}$;

Model 2: $\alpha = 0.9 \cdot 10^{-6}$, $m = \dfrac{1}{3} 10^{-11} eV$, $\gamma_0 = 9 \cdot 10^{20}$;

Model 3: $\alpha = 0.45 \cdot 10^{-6}$, $m = \dfrac{1}{27} 10^{-31} eV$, $\gamma_0 = 81 \cdot 10^{40}$

It is evident that despite the enormous differences between parameters of the models, the actual differences in superluminal anomalies are rather small.

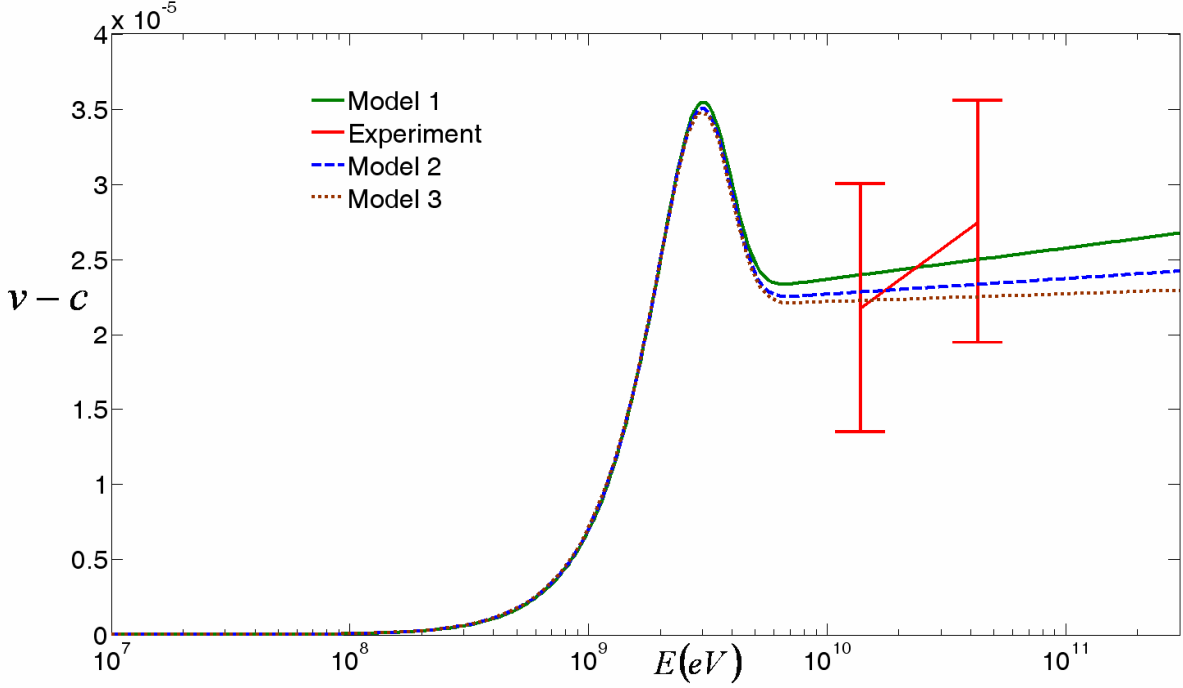

Fig 4. Illustration of approximate invariance of superluminal behavior of high-energy neutrinos during renorm transformations.

This approximate invariance of behavior of high-energy neutrinos during renorm transformations complicates identification of unambiguous dispersion relation for neutrino.

Nevertheless, note that low-energy neutrinos can reveal themselves to great extent in form of dark matter in gravitational large-scale events.

### 2.4. Models of hyperbolic sine, hyperbolic tangent and logarithm

Previously, we discussed a base power model with slow hyper-relativistic adjustment and its modified version. Such types of models have the most solid grounds to aspire to adequately describe possible anomalies related to motion of neutrino. Nevertheless, methodically it is reasonable to cover other elementary hyper-relativistic models. Moreover, we cannot eliminate the possibility that for different types of particles, we may get different dispersion relations.

As previously, $\gamma_0$ is the critical gamma-factor that controls the switch from relativistic to hyper-relativistic mode.

The hyperbolic sine model is described by the following functions:

$$f(E) = m\gamma_0 \sinh\left(\frac{E}{m\gamma_0}\right) \qquad (2.4.1)$$

$$F(E) = 2m^2\gamma_0^2\left[\cosh\left(\frac{E}{m\gamma_0}\right) - \cosh\left(\frac{1}{\gamma_0}\right) + \frac{1}{2\gamma_0^2}\right], \qquad (2.4.2)$$

The hyperbolic tangent model has the following form:

$$f(E) = m\gamma_0 \tanh\left(\frac{E}{m\gamma_0}\right) \qquad (2.4.3)$$

$$F(E) = 2m^2\gamma_0^2\left[\ln\left(\cosh\left(\frac{E}{m\gamma_0}\right)\right) - \ln\left(\cosh\left(\frac{1}{\gamma_0}\right)\right) + \frac{1}{2\gamma_0^2}\right], \qquad (2.4.4)$$

Finally, the logarithm model is defined as:

$$f(E) = m\gamma_0 \ln\left(1 + \frac{E}{m\gamma_0}\right) \tag{2.4.5}$$

$$F(E) = 2m^2\gamma_0^2\left[\left(1 + \frac{E}{m\gamma_0}\right)\left(\ln\left(1 + \frac{E}{m\gamma_0}\right) - 1\right) - \left(1 + \frac{1}{\gamma_0}\right)\left(\ln\left(1 + \frac{1}{\gamma_0}\right) - 1\right) + \frac{1}{2\gamma_0^2}\right] \tag{2.4.6}$$

Note that the hyperbolic sine model does not allow for superluminal motion of particles, while there are no such constraints for hyperbolic tangent and logarithm (Fig.5). It can also be seen that for $\gamma \ll \gamma_0$ the three curves practically coincide.

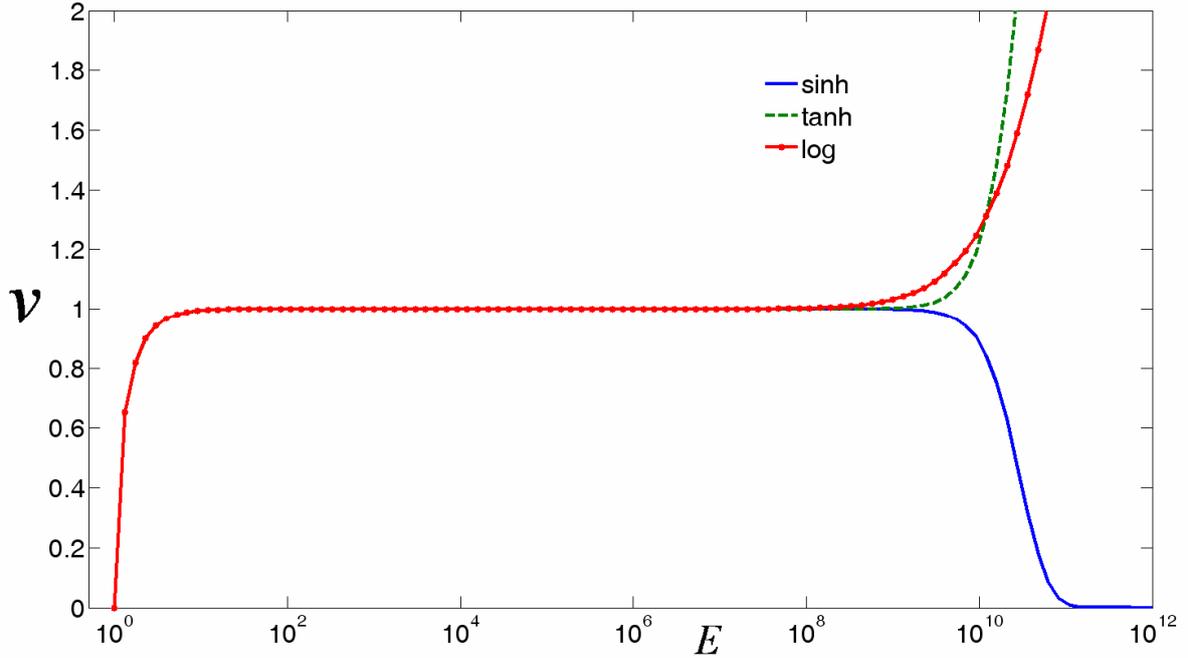

Fig.5. Dependence of velocity on energy for the three elementary hyper-relativistic models ($m = 1$, $\gamma_0 = 1 \cdot 10^{10}$)

## 3. Examples of numerical calculations of hyper-relativistic dynamics of particles
### 3.1. Composition of motion examples

In this section we will use power model with slow hyper-relativistic adjustment (2.1.1). Here and below parameter values are as follows: $\alpha = 2 \cdot 10^{-6}$, $\gamma_0 = 10^{10}$.

In the first and second example the particles have relatively small, yet ultra-relativistic energies. As such, adjustments to standard relativity theory are small.

**Example 1.** Equal weight particles $m_1 = m_2 = 1$ move towards one another with the same energies $E_1 = E_2 = 100$. Respective velocities of particles are:

$$v_1 = -v_2 = 0.999949998700039589786498.$$

Then, by the method described in section 1.5, we can calculate velocity of the second particle in reference frame connected to the first particle:

$$v_2' = -0.999999998791485653195285$$ (minus sign stands for direction). Calculation using standard Lorentz formula (1.5.21) gives a similar result:

$$v_2'^{(L)} = -0.999999987498749994948164$$

In this example initial velocities $v_1$ and $v_2$ in laboratory reference frame are close to those for the standard Lorentz relation $F(E) = E^2$ that are in turn equal to:

$$v_1^{(L)} = -v_2^{(L)} = 0.999949998749937496093477$$

**Example 2.** Here again $m_1 = m_2 = 1$, but the second particle chases the first one: $E_1 = 1000$, $E_2 = 2000$. Respective velocities of particles are:
$v_1 = 0.999999500001356549561746$,
$v_2 = 0.999999875003232547697757$

In reference frame connected to the first particle, velocity of the second particle is:
$$v_2' = 0.600001792663232764608785$$

Once again, calculation using standard Lorentz formula (1.5.21) gives a similar result:
$$v_2'^{(L)} = 0.600007467184247247399510$$

Note that calculations of invariant mass of the system also yields similar results both for hyper-relativistic ($M$) and standard models ($M_0$):

$$M = 2.121321333879268779668$$
$$M_0 = 2.121320409850943705570534$$

The method of calculation of mass of a system in hyper-relativistic theory is given in section 1.4. It can further be shown that in standard relativity theory the mass of a two-particle system is equal to:

$$M_0^2 = m_1^2 + m_2^2 + 2m_1 m_2 \left( \gamma_1 \gamma_2 \pm \sqrt{(\gamma_1^2 - 1)(\gamma_2^2 - 1)} \right) \quad (3.1.1)$$

In this formula the plus sign stands for particles moving toward one another, the minus sign - for one particle chasing the other.

Here the velocities are close to standard Lorentz model again.

$$v_1^{(L)} = 0.999999499999874999937500$$
$$v_2^{(L)} = 0.999999874999992187499023$$

**Example 3.**

This example also corresponds to sub-luminal velocities, but much closer to the speed of light than the standard Lorentz model. We will see that in this case there is a significant modification in the law of composition of velocities.

Let $m_1 = m_2 = 1$, while particle #2 chases particle #1. Also let

$E_1 = 60121.3$, $E_2 = 60121.367$, so that the velocities are very close to the speed of light.

$$v_1 = 0.999999999999999510295504,$$
$$v_2 = 0.999999999999999986161690$$

In the frame connected to the first particle, velocity of the second particle is equal to:

$$v'_2 = 0.16103945824945666995 3155 \cdot 10^{-5}$$

Calculation using standard Lorentz formula (1.5.20) gives a very different result:

$$v'^{(L)}_2 = 0.94503621248172405 0705898$$

Therefore, hyper-relativistic composition of velocities results in practically non-relativistic motion, while standard theory gives a much higher number.

If the standard law of dispersion were valid, velocities of the particles would be equal to:

$$v^{(L)}_1 = 0.99999999986167098678 1125$$
$$v^{(L)}_2 = 0.99999999986167129509 2103$$

In reality, in present example velocities are much closer to the speed of light and as such there are large deviations from the standard theory.

Finally, note that the masses add almost perfectly and the mass is rather small.

$$M = 2.00000000000064834267783 \text{ for hyper-relativistic model}$$
$$M_0 = 2.00000000000031047912423 \text{ for standard model}$$

**Example 4** is identical to the previous example but the particles move towards one another. In standard model relative velocity cannot exceed the speed of light, but in modified framework it does exceed it though by relatively small amount.

Let $m_1 = m_2 = 1$, $E_1 = E_2 = 60121.367$. Then

$$v_1 = -v_2 = 0.999999999999999986161690$$

However, in the reference frame related to the first particle, the absolute velocity of the second particle somewhat exceeds the speed of light:

$$v'_2 = -1.000000601545370757 91808.$$

It is clear that the standard Lorentz law of composition of velocities (1.5.21) cannot yield a velocity exceeding the speed of light. Calculation result confirms this:

$$v'^{(L)}_2 = -0.9999999\ 9999999999\ 9999999999\ 9999999042\ 5058961250\ 1$$

Note that for head-on collision of particles with equal energies, invariant mass is simply equal to total energy

$$M = M_0 = 120242.734$$

In hyper-relativistic theory the change in kinematics compared to Lorentz case can have significant qualitative consequences, as we will illustrate in two more examples below.

**Example 5.**
Here we demonstrate that the total mass of a system can be less than the sum of masses of individual particles. Such phenomenon is impossible in standard relativity theory, but may take place in hyper-relativistic mechanics.

Let $m_1 = m_2 = 1$, $E_1 = 10^{11}$, $E_2 = 2 \cdot 10^{11}$. Then velocities of the particles exceed the speed of light (second particle chases the first one):

$$v_1 = 1.00002632878509901368267,$$
$$v_2 = 1.00002702194878977610802$$

The value of total mass in standard Lorentz theory (3.1.1) exceeds the sum of individual masses:

$$M_0 = 2.12132034355964257320254$$

This is not the case in hyper-relativistic model

$$M = 1.88988174150476143298142$$

In the example, the gamma-factor of the second particle in reference frame connected to the first particle is less than unity

$$\gamma'_2 = 0.78582649843653495100 9017$$

Then, according to the energy-momentum invariant, squared momentum of the second particle becomes negative and the velocity has only imaginary component.

$$p'^2_2 = -0.38247671435497450 8805705$$
$$v'_2 = 0.78700199020820154 9862666 \cdot i$$

Therefore, the two-particle system becomes interconnected (in fact, the two particles practically become one) when we move on from the laboratory reference frame to a frame connected to one of the particles. It means that a particle with mass $M = 1.88988...$ can break into two particles with unitary masses during its motion, but this cannot happen when it is static.

In general, behavior of matter in superluminal systems can seem rather abnormal. Note, however, that though the possibility of construction of such "superluminal laboratories" is obscure, that does not rule out existence of superluminal particles in our present laboratories and related reference frames.

The following example demonstrates how the result of composition may depend on the masses (mass ratio).

**Example 6**
Let our particles move towards one another, while having the same gamma-factors: $\gamma_1 = \gamma_2 = 10^{10}$. Then energies and velocities of the particles are as follows:

$$E_1 = 10^{10} m_1$$
$$E_2 = 10^{10} m_2$$
$$v_1 = -v_2 = 1.00002796865977242124243$$

It appears then that the law of composition of velocities, i.e. velocity of the second particle in reference frame connected to the first particle depends on the mass ratio:

For instance, if $m_1 = m_2 = 1$, then

$$v_2' = -1.000029603869447596200099$$

While if $m_1 = 10$, $m_2 = 1$ then

$$v_2' = -1.000030735827179055575415$$

Similarly, if $m_1 = 1$, $m_2 = 10$ then

$$v_2' = -1.000028472752379089696624$$

### 3.2. Hyper-relativistic Cherenkov radiation

Previously we considered one-dimensional motion. Now let us account for real geometry by considering a problem of Cherenkov radiation of photon and mass particle by a fast-moving particle in vacuum. Such phenomenon is not allowed in standard relativity theory, but is permitted in hyper-relativistic framework.

We assume that a photon with energy $E_\gamma$ is radiated at angle $\theta_\gamma$ to particle trajectory. We further suppose that the usual law of dispersion is valid for the photon $E_\gamma = P_\gamma$. Next, according to the laws of conservation of momentum and energy, angular distribution (dependence of photon energy on its angle of radiation) is as follows:

$$\cos\theta_\gamma = \frac{F(E) + E_\gamma^2 - F(E - E_\gamma)}{2E_\gamma \sqrt{F(E) - m^2}} \qquad (3.2.1)$$

Then, for the standard Lorentz law of dispersion $F(E) = E^2$ we get:

$$\cos\theta_\gamma = \frac{E}{p} = \frac{1}{v}, \qquad (3.2.2)$$

This implies that $|\cos\theta_\gamma| > 1$, which is impossible due to prohibition of superluminal motion in standard theory. Therefore, according to the standard theory, a particle cannot radiate in vacuum.

Next, let us consider hyper-relativistic motion. It is interesting to analyze low-energy (low-frequency) photons, when $E_\gamma \to 0$. Then, if we limit ourselves to first order approximation in Taylor series expansion, we derive $F(E - E_\gamma) \approx F(E) - \frac{\partial F}{\partial E} E_\gamma$.

Afterwards, we shall neglect the second infinitesimal order summand $E_\gamma^2$ and take (1.2.10) into account to obtain the classic Cherenkov condition again:

$$\cos\theta_\gamma = \frac{1}{v} \qquad (3.2.3)$$

Now, however, this condition is feasible, since velocity of particle can exceed the speed of light in vacuum.

Next, consider Cherenkov radiation of a non-zero mass particle. Assume that a particle with mass $m$ and energy $E$ radiates another particle with mass $m_2$ and energy $E_2$ at angle $\theta_2$. Here $m_1 = m$, as the original particle does not change.

In that case we shall obtain a more general formula instead of (3.2.1):

$$\cos\theta_2 = \frac{p^2 - p_1^2 + p_2^2}{2pp_2} \qquad (3.2.4),$$

where $p$, $p_1$, $p_2$ - are absolute values of momenta of particles $m$, $m_1$ и $m_2$.

We assume that the radiated particle follows the ordinary law of dispersion. Then the equation above can be rewritten as

$$\cos\theta_2 = \frac{F(E) - F(E - E_2) + E_2^2 - m_2^2}{2\sqrt{F(E) - m^2}\sqrt{E_2^2 - m_2^2}} \qquad (3.2.5)$$

Figures 6 and 7 show angular distributions for Cherenkov radiation of a photon and a particle with mass $m_2 = 1 MeV$. The latter case may serve as a raw estimate of electron-positron pair radiation (when there is no mutual motion of electron and positron). We consider a modified power model (2.2.3) with parameters: $k = 2$; $\alpha = 2 \cdot 10^{-6}$, $\gamma = 10^{11}$. Neutrino mass is assumed to be equal to $0.1 eV$. Spectra of neutrinos with energy values 10GeV, 5GeV, 2Gev, 1.3GeV are presented.

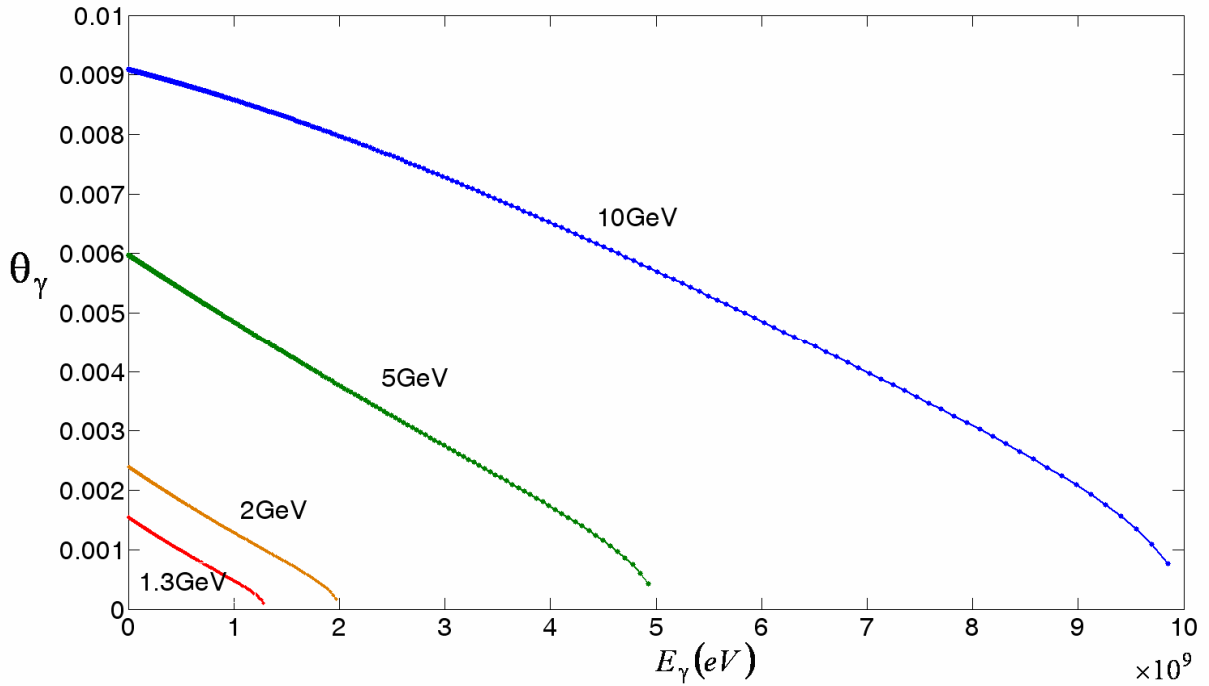

Fig 6. Angular distribution for Cherenkov photon radiation

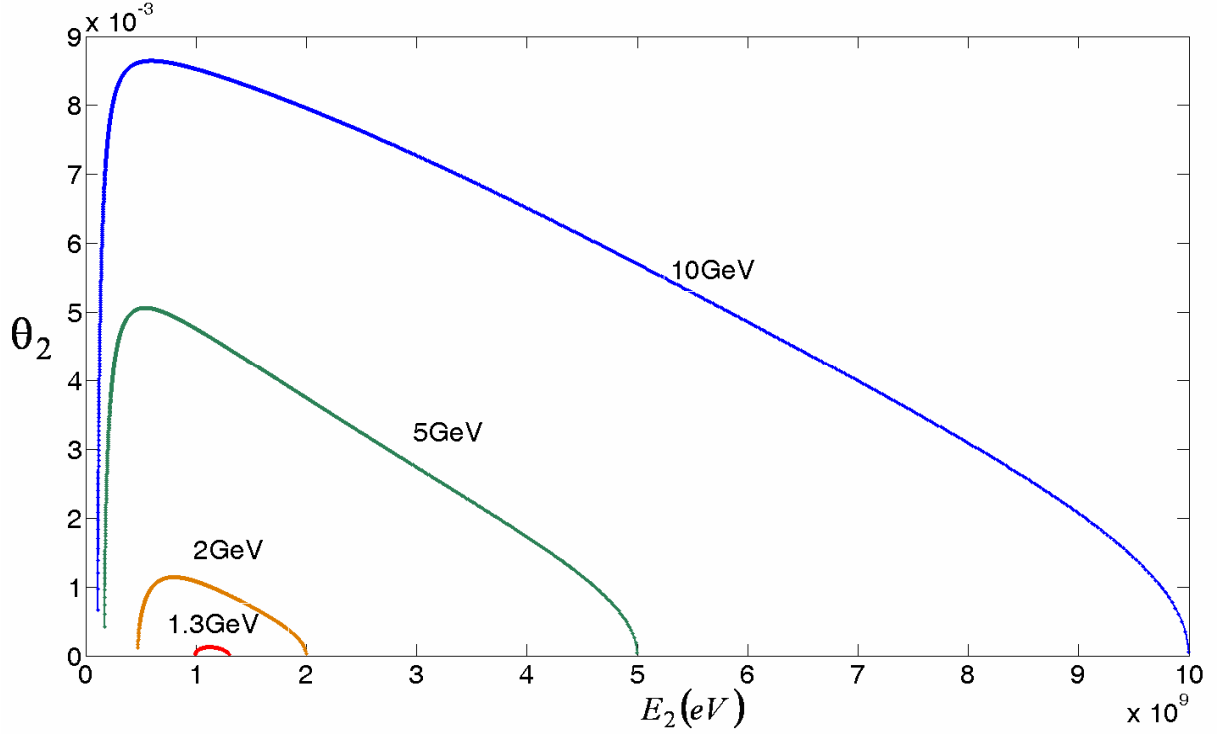

Fig 7. Angular distribution for Cherenkov radiation of particle with mass 1 MeV.

Note that in our case the threshold for creation of the electron-positron pair is around 1.2 GeV, which is significantly larger than the 140 MeV estimate in [9] by Cohen and Glashow. We believe that this is explained by incorrect disregard of dependence of neutrino's superluminal velocity on its energy,

Authors of [10] did attempt to account for this dependence by considering a model that assumes that superluminal velocity of a neutrino is in quadratic relationship to its energy. Note, however, that such model is not consistent with experimental data, as according to [1] superluminal anomaly is almost the same for energies 13.9 GeV and 42.9 GeV (also shown on Fig.1). At the same time, the quadratic model would imply 9.5 times larger anomaly for 42.9 GeV compared to 13.9 GeV.

Also, note that according to [10] the impact of Cherenkov electron-positron radiation on neutrinos in the OPERA experiment is small, while according to [9] it has much larger effect.

Finally, we highlight that, in our view, the modified power model with slow hyper-relativistic adjustment (2.2.3) provides the best fit for available data on superluminal anomalies for a wide band of energies.

**4. Hyper-relativistic Dirac equation**
**4.1. Application of Dirac's approach to hyper-relativistic dispersion relation.**

When considering a quantum framework, we need to replace energy with Hamiltonian ($E \to H$). Next, in order to derive a hyper-relativistic Dirac equation and following (1.3.10) we consider function $G(H)$ such as:

$$G(H) = \sqrt{F(H)} = \sqrt{p^2 + m^2} \qquad (4.1.1)$$

For standard Lorentz model $G(H) = H$, but this is not true in general. According to Dirac's approach:

$$\sqrt{p^2 + m^2} = \vec{\alpha}\vec{p} + \beta m \tag{4.1.2}$$

Here $\vec{\alpha}$ and $\beta$ are Dirac matrices that have the following form:

$$\vec{\alpha} = \begin{pmatrix} 0 & \vec{\sigma} \\ \vec{\sigma} & 0 \end{pmatrix}, \beta = \begin{pmatrix} I & 0 \\ 0 & -I \end{pmatrix} \tag{4.1.3}$$

Therefore:

$$G(H) = \vec{\alpha}\vec{p} + \beta m \tag{4.1.4}$$

For any particular case, we can derive function $G(H)$ that corresponds to the model. Hamiltonian $H$ that is a function of momentum operator $\vec{p}$ is the root of equation (4.1.4).

Non-stationary hyper-relativistic Dirac equation is similar to Schrodinger equation:

$$i\frac{\partial \psi}{\partial t} = H\psi \tag{4.1.5}$$

It is important to note, though, that now $H \neq \vec{\alpha}\vec{p} + \beta m$, because $H$ is a solution to equation (4.1.4).

Stationary hyper-relativistic Dirac equation has the form:

$$G(H)\psi = (\vec{\alpha}\vec{p} + \beta m)\psi \tag{4.1.6}$$

The wave function $\psi$ is a four-component spinor (bispinor).

$G(H)$ can be rather complex. Yet, let us suppose that it can be decomposed into a well-defined and convergent series:

$$G(H) = g_0 + g_1 H + g_2 H^2 + \ldots \tag{4.1.7}$$

In a stationary state $H\psi = E\psi$, $H^2\psi = E^2\psi$ etc., therefore:

$$G(H)\psi = (g_0 + g_1 H + g_2 H^2 + \ldots)\psi = \\ = (g_0 + g_1 E + g_2 E^2 + \ldots)\psi = G(E)\psi \tag{4.1.8}$$

**4.2. Free hyper-relativistic motion of Dirac particle**

We shall consider free motion with defined energy $E$ and momentum $\vec{p}$ as a particular example. Such motion will be described by some parameter $G = G(E)$.

Let $\psi = \begin{pmatrix} \varphi \\ \chi \end{pmatrix}$ be a bispinor of free motion that we need to calculate.

Then according to stationary hyper-relativistic equation (4.1.6), we get a system of two homogeneous equations

$$(G - m)\varphi - \vec{\sigma}\vec{p}\chi = 0 \tag{4.2.1}$$

$$-\vec{\sigma}\vec{p}\varphi + (G+m)\chi = 0 \qquad (4.2.2)$$

For the system to be solvable its determinant has to be equal to zero:

$$\det\begin{pmatrix} G-m & -\vec{\sigma}\vec{p} \\ -\vec{\sigma}\vec{p} & G+m \end{pmatrix} = 0, \qquad (4.2.3)$$

therefore: $G^2 = p^2 + m^2$ \qquad (4.2.4)

We see that the situation is similar to the ordinary Dirac equation, where energy $E$ is replaced by parameter $G = G(E)$.

From (4.2.2) one can derive the relation between spinors $\varphi$ and $\chi$

$$\chi = \frac{\vec{\sigma}\vec{p}}{(G+m)}\varphi \qquad (4.2.5)$$

**4.3. Effect of external electromagnetic field**

When an external electromagnetic field is applied to a charged Dirac particle, we modify the variables as follows:

$$\vec{p} \to \vec{p} - e\vec{A} \qquad (4.3.1)$$

$$H \to H - eA_0 \qquad (4.3.2)$$

Here $A_0$ and $\vec{A}$ are scalar and vector potentials, $e$ is the charge of the particle.

Let $G(H) = H + \delta G$, \qquad (4.3.3)

where $\delta G$ is a small perturbation.

If we rewrite this equation as $H = G - \delta G$, we may interpret value $-\delta G$ as a perturbation to Hamiltonian.

Approximate adjustments to energy states of a system can be found by stationary perturbation theory. At the same time it suffices to know wave function to zero approximation to calculate adjustment to energy.

Let $H\psi = E\psi$, \qquad (4.3.4)

where $E = E_0 + \delta E$ \qquad (4.3.5).

Here $E_0$ is the unperturbed energy, $\delta E$ - perturbation.

The unperturbed state of the system is a solution to ordinary Dirac equation:

$$\left[\vec{\alpha}(\vec{p}-e\vec{A}) + \beta m + eA_0\right]\psi_0 = E_0\psi_0 \qquad (4.3.6)$$

Perturbation $\delta E$ to energy of the system is defined by the following matrix element:

$$\delta E = -\langle\psi_0|\delta G|\psi_0\rangle \qquad (4.3.7)$$

Here $\delta G$ is a function of $E_0 - eA_0$

## 5. Conclusion

Let us briefly formulate the main results of this paper.

First of all, we attempt to propose a new hyper-relativistic form of mechanics that is based on modified dispersion relations and energy-momentum invariants. Classical Newtonian mechanics and special relativity theory are particular limit cases of the new framework. It appears that predictions of hyper-relativistic theory significantly differ from standard relativistic mechanics, so that the Lorentz gamma factor is very large.

Secondly, we formulate the principle of hyper-relativity that generalizes the usual relativity principle and yields a radical change in the law of composition of velocities and particle kinematics in general. We apply the new framework to discuss models that may adequately explain hypothetical superluminal motion of high-energy neutrinos as well as Cherenkov radiation of photon and non-zero mass particle in vacuum.

Finally, we derive a hyper-relativistic Dirac equation and study free motion of a particle as well as impact of external electromagnetic field.

Authors would like to thank prof. Yu.G. Rudoi for a very insightful discussion.